\documentclass[a4paper,fleqn,usenatbib]{mnras}

\usepackage{newtxtext,newtxmath}

\usepackage[T1]{fontenc}
\usepackage{ae,aecompl}

\usepackage{graphicx}	
\usepackage{amsmath}	
\usepackage{amssymb}	
\usepackage{color}

\title[Cold Spot: masking and the expected ISW]{The cosmic microwave background Cold Spot anomaly: the impact of sky masking and the expected contribution from the Integrated Sachs-Wolfe effect}

\author[K. Naidoo, A. Benoit-L\'{e}vy and O. Lahav]{
	Krishna Naidoo,$^{{\color{blue}1\star}}$ Aur\'{e}lien Benoit-L\'{e}vy$^{{\color{blue}2\star}}$ and Ofer Lahav$^{{\color{blue}1}}$\thanks{E-mail: \href{mailto:krishna.naidoo.11@ucl.ac.uk}{krishna.naidoo.11@ucl.ac.uk} (KN); \href{mailto:benoitl@iap.fr}{benoitl@iap.fr} (AB-L); \href{mailto:o.lahav@ucl.ac.uk}{o.lahav@ucl.ac.uk} (OL)}
	\\
	$^{1}$Department of Physics \& Astronomy, University College London, Gower Street, London WC1E 6BT, UK\\
	$^{2}$Sorbonne Universit{\'e}s, UPMC Univ Paris 6 et CNRS, UMR 7095, Institut d'Astrophysique de Paris, 98 bis bd Arago, F-75014 Paris, France\\
}

\date{Accepted XXX. Received YYY; in original form ZZZ}

\pubyear{2017}

\begin{document}
\label{firstpage}
\pagerange{\pageref{firstpage}--\pageref{lastpage}}
\maketitle

\begin{abstract}
We re-analyse the cosmic microwave background (CMB) Cold Spot (CS) anomaly with particular focus on understanding the bias a mask (contaminated by Galactic and point sources) may introduce. We measure the coldest spot, found by applying the Spherical Mexican Hat Wavelet transform on 100,000 cut-sky (masked) and full-sky CMB simulated maps. The CS itself is barely affected by the mask; we estimate a 94 per cent probability that the CS is the full-sky temperature minimum. However, $\sim$48 per cent (masked fraction of the mask) of full-sky minima are obscured by the mask. Since the observed minima are slightly hotter than the full-sky ensemble of minima, a cut-sky analysis would have found the CS to be significant at $\sim$2.2$\sigma$ with a wavelet angular scale of $R=5^{\circ}$. None the less, comparisons to full-sky minima show the CS significance to be only $\sim$1.9$\sigma$ and $<$2$\sigma$ for all R. The CS on the last scattering surface may be hotter due to the Integrated Sachs-Wolfe effect in the line of sight. However, our simulations show that this is on average only $\sim$10 per cent (about $−10\mu K$ but consistent with zero) of the CS temperature profile. This is consistent with Lambda and Cold Dark Matter reconstructions of this effect based on observed line-of-sight voids.
\end{abstract}




\section{Introduction}

The cosmic microwave background (CMB) Cold Spot (CS) anomaly was discovered by \citet{Vielva2004} using the Spherical Mexican Hat Wavelet (SMHW) \citep{Cayon2001} on WMAP data. The anomaly has persisted \citep{Wmap2007,Wmap2011} and was later verified by Planck \citep{PlanckIso2015}.

\citet{Inoue12007,Inoue22007} claimed the Integrated Sachs-Wolfe (ISW) \citep{Sachs1967} and Rees-Sciama (RS) \citep{Rees1968} effects of a large void at redshift $z\sim1$ could explain the entire feature (\citet{Nadathur2014} show the RS is subdominant in all cases). However, pencil beam surveys \citep{Bremer2010,Granett2010} have effectively ruled out the possibility of such a large void at high redshift (i.e. $0.5<z<1$). Studies of the galaxy distribution in the relevant region using photo-z initially appeared to indicate that a single spherical/elliptical void exists along the line-of-sight (LOS) at lower redshift \citep[see][]{Szapudi2015,AndrasJuan2015}. Several studies have shown this is insufficient to explain the CS \citep[see][]{Nadathur2014,Zibin2014,Marcos2016}. \cite{Naidoo2016} found that a model using multiple voids could only explain a fraction of the feature. This was recently confirmed by \citet{Mackenzie2017} who observed three voids along the LOS and came to the same conclusion. Hints of a stronger than expected ISW signal have been found in some stacked void studies \citep{Granett2008,Cai2014,DES2017,Kovacs2017}, leading to speculation that the causal relation between the CS and the LOS voids may be much greater than that predicted by the ISW. However, \citet{Ilic2013}, \citet{Hotchkiss2015} and \citet{NadaCrit2016} have found no such excess and obtain results consistent with $\Lambda$CDM.

The use of a mask in the SMHW analysis of the CS, to minimise contribution from the Galaxy and point sources, is common practice \citep[see][]{Vielva2004,Zhang2010,Nadathur2014,PlanckIso2015}. Because the SMHW transform integrates across the sky, contributions from masked areas will leak to neighbouring regions. Thus a more aggressive mask than the original is applied to the filtered map \citep[see][]{Zhang2010,Rassat2014}. While the application of a mask is sometimes unavoidable, \citet{Rassat2014} show that many CMB anomalies, including the CS, are no longer significant when carried out without the use of a mask on full sky LGMCA CMB maps \citep{Bobin2014}. Furthermore, the CS's inability to be detected by other filters \citep[see][]{Zhang2010,Marcos2017} has placed doubt on its significance. However, this is often argued to be due to the SMHW sensitivity to what makes the CS anomalous, i.e. its high transition from cold to hot.

In this paper, we investigate the effects of masking on the detection and resulting significance of the CS and the expected contribution of the ISW to the CS profile.

\section{Method}

In the following analysis we use the Planck SMICA CMB map and the Planck Common Field mask\footnote{Available from \href{http://pla.esac.esa.int/pla/\#home}{http://pla.esac.esa.int/pla/\#home}.}.

\subsection{Spherical Mexican Hat Wavelet}

The Spherical Mexican Hat Wavelet (SMHW) is defined according to an angular scale $R$ as:
\begin{equation}
	\Psi(\theta;R) = A_{wav}(R)\left(1+\left(\frac{y}{2}\right)^{2}\right)^{2}\left(2-\left(\frac{y}{R}\right)^{2}\right)\exp\left(-\frac{y^{2}}{2R^{2}}\right),
\end{equation}
where $y\equiv 2\tan(\theta/2)$ and $\theta$ is the angular separation between two points, $\hat{n}$ and $\hat{n}'$, on a sphere. $A_{wav}(R)$ is a normalisation constant defined as:
\begin{equation}
	A_{wav}(R) = \left[2\pi R^{2}\left(1+\frac{R^{2}}{2}+\frac{R^{4}}{4}\right)\right]^{-1/2}.
\end{equation}

The filtered temperature, i.e. the SMHW value of a point at $\hat{n}$ as the transform is applied to an area with an angular radius of $\theta$, is given by:
\begin{equation}
\label{deltat}
	\Delta T_{wav}(\theta;\hat{n},R) = \int_{0}^{\theta}\Delta T(\vec{n}')\Psi(\theta';R)d\Omega',
\end{equation}
where $\hat{n}'$ are pixels located within an angular distance $<\theta$ from point $\hat{n}$. Such pixels are found by using the \texttt{HEALPix} function \texttt{query\_disc}. The SMHW of a single pixel, $\Delta \mathcal{T}_{\Psi}(\hat{n})$, is then calculated by integrating equation \ref{deltat} across the whole sky or up to an angular radius of $\theta \simeq 4R$ (since $\Psi\left(\theta\gtrsim4R;R\right)\simeq0$):
\begin{equation}
\label{smhw}
	\Delta \mathcal{T}_{\Psi}(\hat{n}) = \Delta T_{wav}(\pi; \hat{n},R) \simeq
	 \Delta T_{wav} (4R; \hat{n},R).
\end{equation}

In order to remove contamination from Galactic foregrounds and point sources a mask is applied. In order to do this we must first calculate an occupancy fraction \citep{Zhang2010}, which determines the contribution of masked regions to the wavelet transform. This is given approximately by:
\begin{equation}
	\mathcal{N}(\hat{n};R) \simeq \int_{0}^{4R} \mathcal{M}(\hat{n}')\Psi^{2}(\theta';R)d\Omega',
\end{equation}
where $\mathcal{M}(\hat{n})$ and $\mathcal{N}(\hat{n})$ are the mask and occupancy fraction value, respectively, at a point $\hat{n}$. Similarly to equation \ref{smhw}, we integrate only up to $\theta=4R$ rather than $\theta=\pi$ for the exact solution since $\Psi\left(\theta\gtrsim4R;R\right)\simeq0$.

\begin{figure}
	\centering
	\includegraphics[width=\columnwidth]{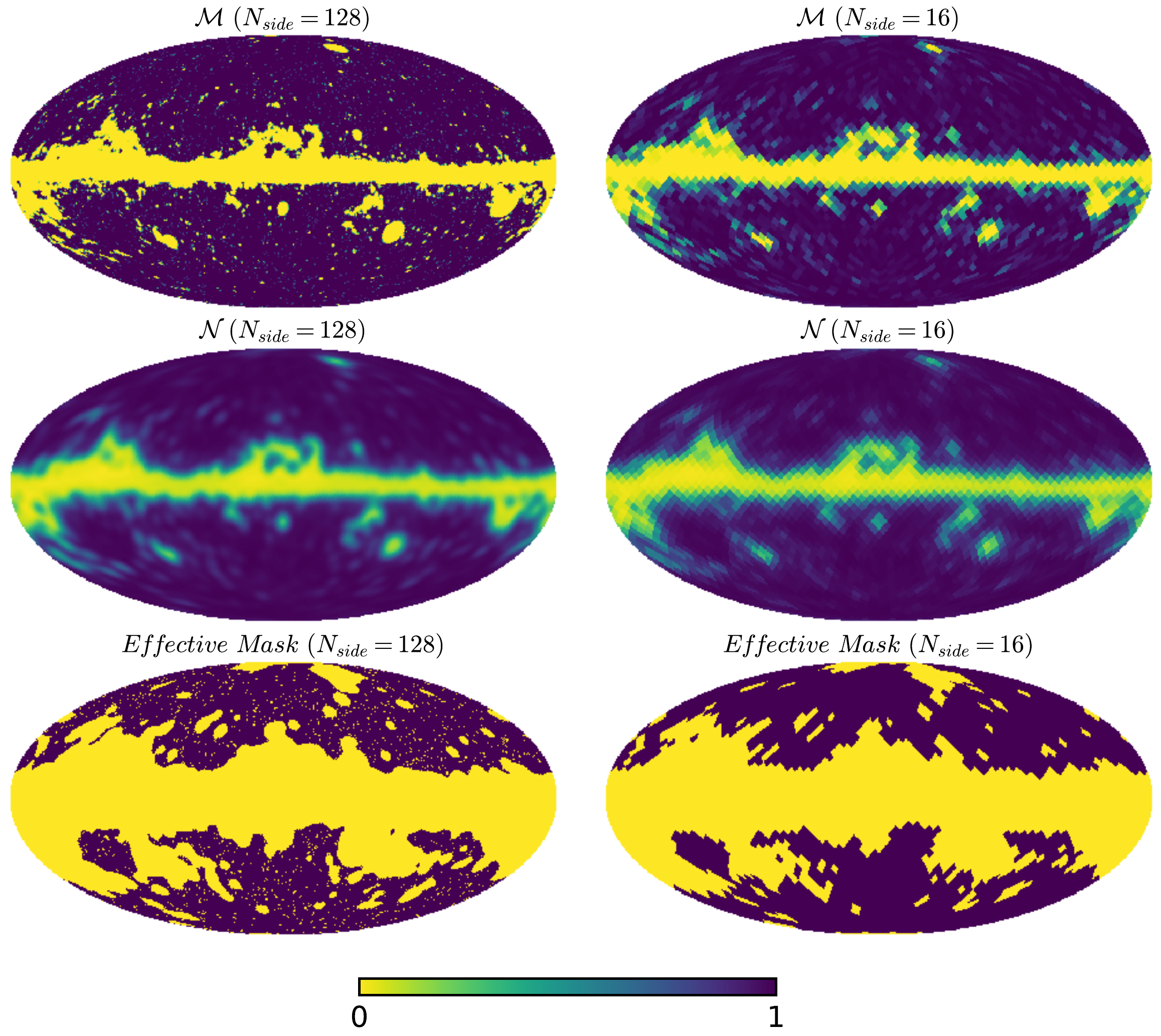}
	\caption{Top panel: the Planck Common Field mask ($\mathcal{M}$), middle panel: the derived occupancy fraction ($\mathcal{N}$) and bottom panel: the effective mask. These are shown at $N_{side}=128$ on the \emph{left} and $16$ on the \emph{right}.}
	\label{mask}
\end{figure}

The SMHW is applied to the full CMB map. Pixels with a mask and occupancy fraction of $\mathcal{M} < 0.9$ or $\mathcal{N} < 0.95$ respectively are then masked to remove areas of the map where contaminated sources may contribute significantly to the result. This means the effective mask applied to the map is considerably larger than the mask $\mathcal{M}$, with $\sim 48$ per cent ($\sim 66$ per cent for  $\mathcal{M} > 0.9$) unmasked pixels (see Fig. \ref{mask}).

\subsection{Simulating Cosmic Microwave Background and Integrated Sachs-Wolfe maps}
Using \texttt{class} \citep{CLASS}\footnote{Software is available from \href{http://class-code.net/}{http://class-code.net/}.} we generate $C_{\ell}$ based on best fit Planck \emph{TT,TE,EE}+\emph{lowP}+\emph{lensing}+\emph{ext} cosmological parameters \citep[see][]{PlanckPara2015}. We deliberately turn off the late ISW effect (i.e. $z<10$), giving $C_{\mathcal{\ell}}$ for the primordial CMB. $C_{\ell}$ for only the late ISW effect are calculated seperately. We then generate primordial CMB maps, $\Delta T_{P}$, and ISW maps, $\Delta T_{ISW}$, using the \texttt{healpix} software \citep{Healpix2005} at $N_{side}=128$ and add them,
\begin{equation}
	\Delta T(\hat{n}) = \Delta T_{P} (\hat{n}) + \Delta T_{ISW} (\hat{n}),
\end{equation}
to give a full CMB map ($\Delta T$). The motivation for generating these maps separately is to allow us to investigate the ISW contribution to the coldest spots in CMB realisations. Since the major contribution to the $\Delta T_{ISW}$ occurs at $z<1.4$ the correlation between $\Delta T_{P}$ and $\Delta T_{ISW}$ is expected to be small.

\subsection{Searching for the coldest spots}

To search for the coldest spots in our simulated maps we apply the SMHW transform to $\Delta T$ maps downgraded from $N_{side}=128$ to $16$. This is carried out with and without a mask. Using the location of the coldest pixel in the downgraded map ($N_{side}=16$) we measure $\Delta T_{wav}(\theta;R)$ (where $R=5^{\circ}$), $\Delta T(\theta)$ and $\Delta T_{ISW}(\theta)$ (i.e. the average $\Delta T_{i}$ of $i$ in concentric rings of the coldest spot) on the original $N_{side}=128$ map. This was carried out on 100,000 simulations. We will refer to the coldest spots identified in unmasked and masked maps as full-sky and cut-sky minima respectively.

To understand the role of masking we additionally measure the angular separation $\alpha$ between the full-sky and cut-sky minima. The two are only considered to be equivalent if $\alpha = 0$, since even a slight misalignment will introduce a bias. We apply the exact same procedure to the Planck SMICA map using the Planck Common Field mask. 

A Frequentist, rather than a Bayesian, approach is applied as we are determining the CS consistency with $\Lambda$CDM rather than doing model comparisons where the alternative would be better suited.

\section{Results}
\subsection{Masked vs unmasked coldest spot}

The full-sky and cut-sky minima are compared in Fig. \ref{obsvstrue}. Using the Planck Common Field mask, we find that these are equivalent only $\sim 48$ per cent of the time, as one would expect given that this is the effective fraction of the map that is removed by the mask. Since cut-sky minima are not always equal to the full-sky minima the use of a mask biases $\Delta \mathcal{T}_{\Psi}$, causing it to be on average $\sim +0.93\mu K$ hotter using the Common Field mask. This is because cut-sky minima are on average $\sim +1.78\mu K$ hotter than the full-sky minima. Interestingly, colder cut-sky minima (i.e. $\Delta \mathcal{T}_{\Psi} < - 18 \mu K$) are more likely to be equivalent to the full-sky minima. This becomes particularly interesting for the CMB CS.

\subsection{The Cold Spot in Planck data}

\begin{figure}
	\centering
	\includegraphics[width=\columnwidth]{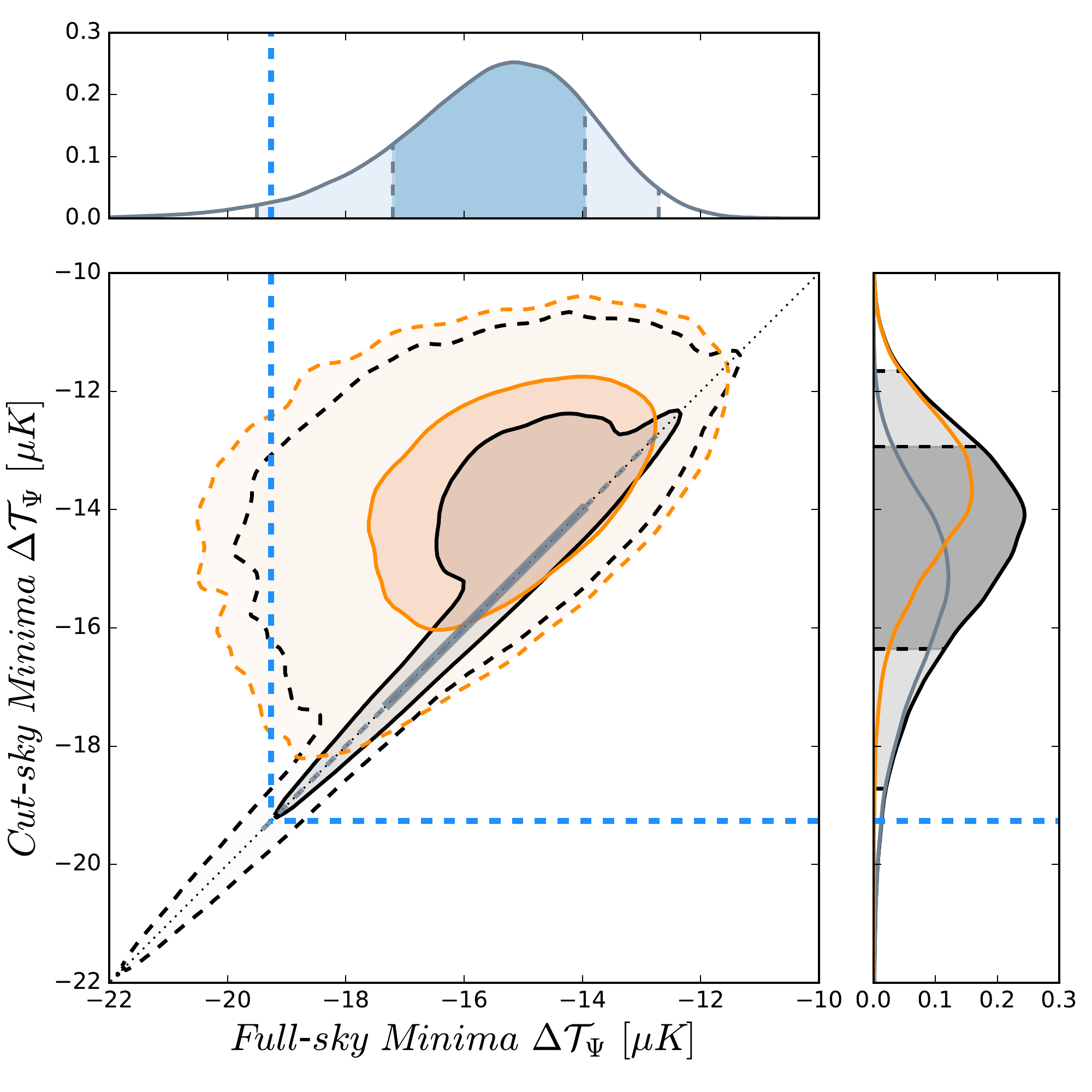}
	\caption{The relation between full-sky and cut-sky minima $\Delta \mathcal{T}_{\Psi}$ are shown. Cases where the two are equivalent are indicated in grey, whilst cases of inequivalence are indicated in orange. Both cases are shown in black. The solid lines and darker shaded contours indicate the $68$ per cent regions and the dashed lines and lighter shaded contours indicate the $95$ per cent regions. The \emph{right} panel shows a kernel density plot of the cut-sky minima. In the \emph{top} panel a kernel density plot of the full-sky minima is shown. Cut-sky minima are shown to be heavily biased due to obscuration of full-sky minima by the mask. This is most prominent for cut-sky minima with $\Delta \mathcal{T}_{\Psi} > -18 \mu K$, since below this it is rare to find cut-sky minima which are not equivalent to the full-sky minima. The CS's $\Delta \mathcal{T}_{\Psi}$ (blue dashed line) is shown for comparison to the cut-sky and full-sky distribution.}
	\label{obsvstrue}
\end{figure}

\begin{figure*}
	\centering
	\includegraphics[width=\textwidth]{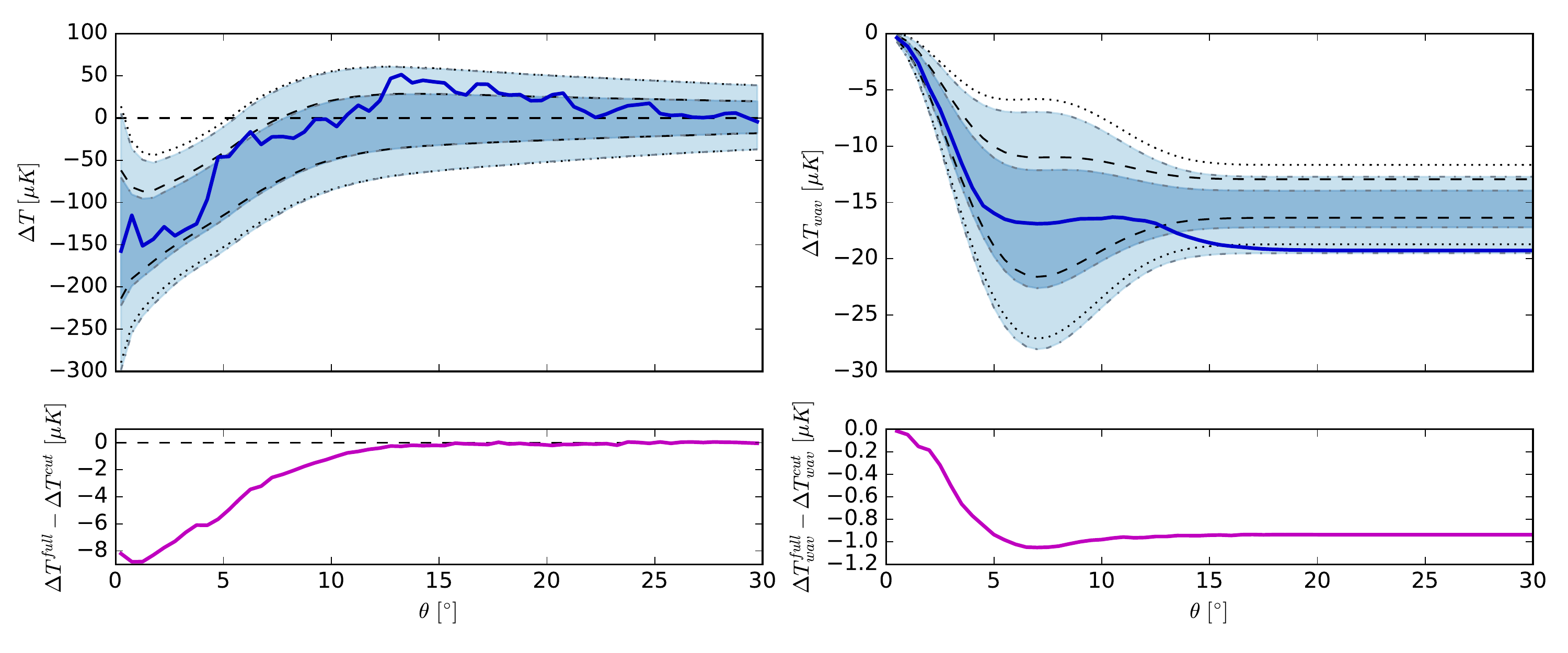}
	\caption{The $1$ and $2\sigma$ contours (dark and light shades respectively) for the $\Delta T(\theta)$ (\emph{top left}) (the average $\Delta T$ in concentric rings from the cut/full-sky minima's center) and $\Delta T_{wav}(\theta)$ (\emph{top right}) profiles, are shown in blue for cut-sky minima in 100,000 simulations. The $1$ and $2\sigma$ contours for cut-sky minima are marked as dashed and dotted black lines, respectively. The CS's $\Delta T(\theta)$ and $\Delta T_{wav}(\theta)$ are shown (measured on Planck's SMICA map) as the dark blue dashed line. The subtle shift in the full-sky $\Delta T(\theta)$ profile around $\theta<5^{\circ}$ shown on the \emph{left} panel appears to lead to colder final temperatures shown on the \emph{right} panel. The difference between the mean of the full-sky and cut-sky $\Delta T$ (\emph{left}) and $\Delta T_{wav}$ (\emph{right}) profiles are indicated with a superscript $full$ and $cut$, respectively, in the bottom panels (note the scale on the bottom panels).}
	\label{cstempprof}
\end{figure*}

The CS has a $\Delta \mathcal{T}_{\Psi} \simeq -19.3\mu K$ with a significance of $\sim 2.2\sigma$ when masked. To make a comparison between the full-sky minima in simulations we must first understand whether the CS is indeed our CMB's full-sky minima. Without any prior knowledge of the CS's $\Delta \mathcal{T}_{\Psi}$ the probability that the cut-sky minima is equivalent to the full-sky minima ($\mathbb{P}(full)$) is $\simeq 0.48$. However, the probability increases as $\Delta \mathcal{T}_{\Psi}$ decreases. The conditional probability that a cut-sky minima similar to the CS (i.e. $-19.5 \mu K< \Delta \mathcal{T}_{\Psi} < -19\mu K$) is equivalent to the full-sky minima ($\mathbb{P}(full|\Delta \mathcal{T}_{\Psi}^{CS})$) is actually $\simeq 0.94$. This means we can be fairly certain that the CS is the CMB's full-sky minima. In Fig. \ref{obsvstrue} the CS's $\Delta\mathcal{T}_{\Psi}$ is shown and lies well within the $2\sigma$ distribution of full-sky minima in simulations. The CS's significance in comparison to full-sky minima is $\sim1.9\sigma$ (which corresponds to a P-value $\sim 3$ per cent). In Fig. \ref{cstempprof} the CS's $\Delta T(\theta)$ and $\Delta T_{wav}(\theta)$ are compared to the $1$ and $2\sigma$ contours of the cut-sky and full-sky minima in simulations (indicated by black lines and blue contours respectively). The comparison illustrates precisely how the observed profiles are biased. For $\Delta T(\theta)$ the main difference occurs near the center ($\theta < 5^{\circ}$) where full-sky minima appear slightly colder. This appears to be more pronounced in $\Delta T_{wav}(\theta)$, where the distribution is found to be consistently colder for all values of $\theta$.

\subsection{The Cold Spot's significance vs. mask size}

\begin{figure}
	\centering
	\includegraphics[width=\columnwidth]{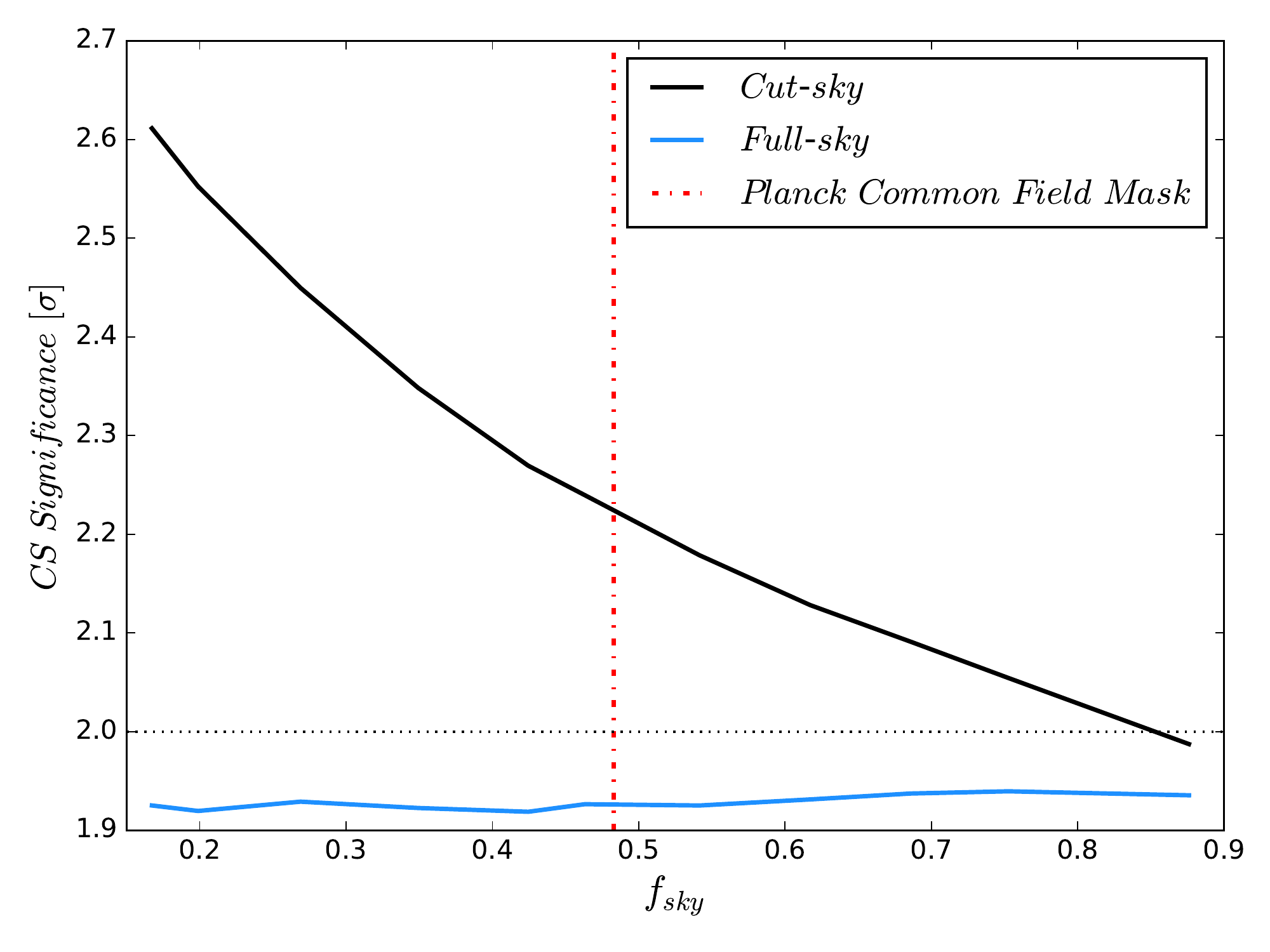}
	\caption{The significance of the CS is measured in comparison to the distribution of cut-sky minima (shown in black) as a function of the mask size ($f_{sky}=$ unmasked fraction of the sky). The significance of the CS is shown in blue in comparison to the full-sky minima observed in a cut-sky. As $f_{sky}$ decreases it is more likely that the full-sky minima is obscured by the mask and that the cut-sky minimum measured is hotter. Consequently these two effects increase the significance of the CS. The vertical red dash-dotted line indicates the $f_{sky}$ of the Planck Common Field mask.}
	\label{maskvsfsky}
\end{figure}

Using the SILC CMB map \cite[][specifically using the $N=5$ map]{Rogers2016} and corresponding mask we test the effect of the size of the mask on the CS's significance. The mask for the SILC CMB map is relatively small such that even the effective mask has $\sim88$ per cent unmasked pixels ($f_{sky}$). We gradually enlarge this mask by masking away a wider Galactic strip and run the same procedure. In Fig. \ref{maskvsfsky} we plot the CS's significance in comparison to cut-sky minima (shown in black) and compare the CS's significance to the full-sky minima (shown in blue) as a function of $f_{sky}$. The CS significance in comparison to cases where the full-sky and cut-sky minima are equivalent always remains $< 2\sigma$. But in comparison to cut-sky minima the significance becomes larger as $f_{sky}$ decreases. Rather unsurprisingly, a larger mask will make it harder to find the full-sky minima and will also make it more likely that a hotter cut-sky minima is measured. The net effect is that a full-sky minima measured in a cut-sky analysis will have a boosted significance due to the size of the mask. This appears to be the case for the CS.

\subsection{The Integrated Sachs-Wolfe for the coldest spots}

\begin{figure}
	\centering
	\includegraphics[width=\columnwidth]{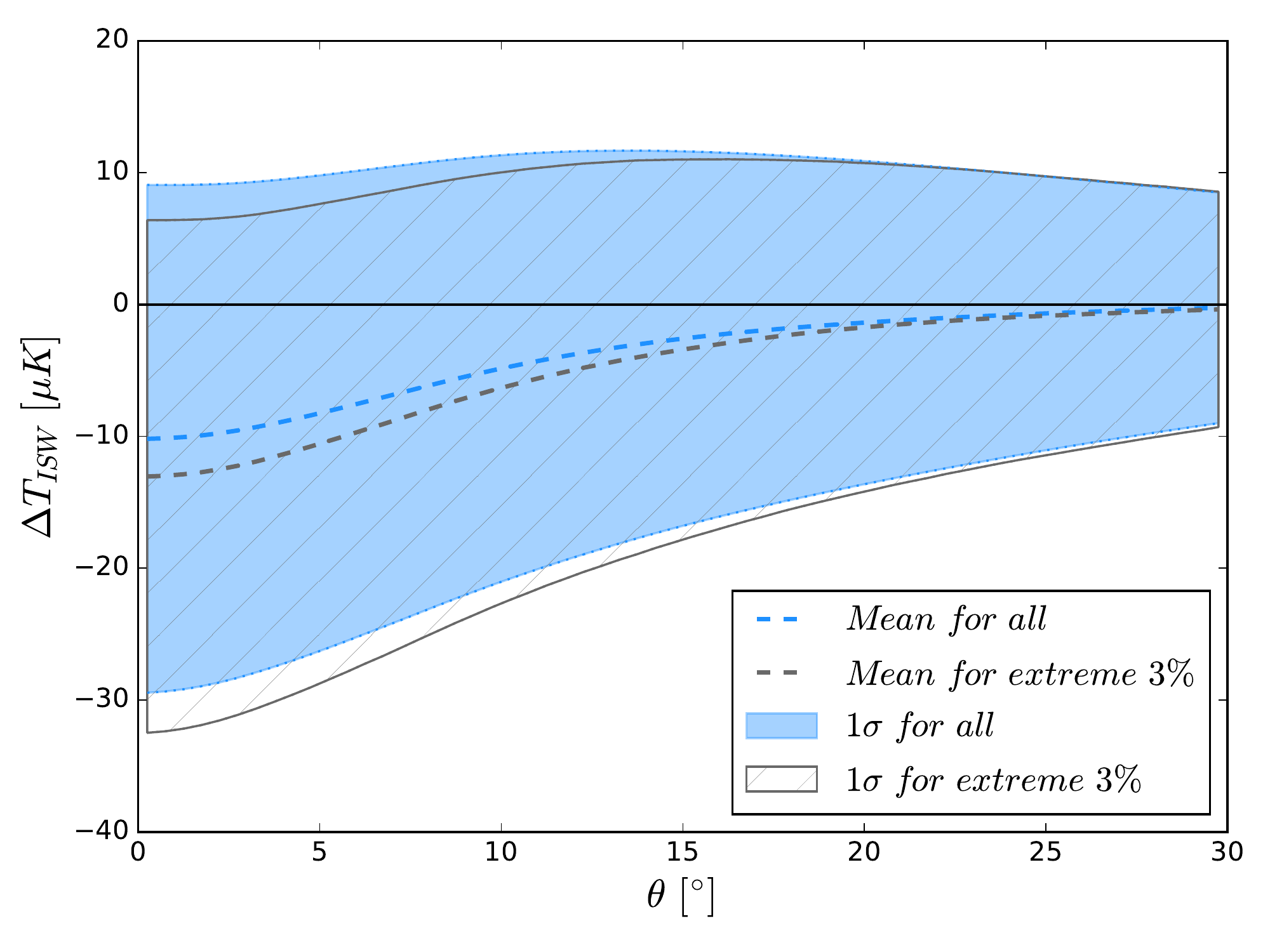}
	\caption{The $1\sigma$ contours of the ISW of all the full-sky minima are shown in blue. The most extreme $3$ per cent are indicated by the grey hatched area. The mean for all the coldest spots and the most extreme $3$ per cent are indicated by the blue and grey dashed lines, respectively. The ISW in either case is not very well constrained and consistent with zero but on average appears to contribute $\sim 10$ per cent to the minima's profiles.}
	\label{isw}
\end{figure}

The ISW contribution to the coldest spots in simulations was measured and is shown in Fig. \ref{isw}. Here we display the mean and $1\sigma$ contours for all the full-sky minima and the most extreme $3$ per cent (which approximately corresponds to the CS's p-value). The profiles are poorly constrained and very similar, with the more extreme case tending to be slightly more negative. The result illustrates that it is very likely that the ISW plays a minor role in the CS profile: $\sim10$ per cent of the full profile. The reconstructed ISW profiles \citep{Rassat2014,Nadathur2014,Finelli2015,PlanckISW2016} appear to be consistent with the predicted ISW shown in Fig. \ref{isw}. The presence of prominant voids in the LOS \citep[see][]{Szapudi2015,AndrasJuan2015} are therefore precisely what we would expect from $\Lambda$CDM.
\begin{table}
	\centering
	\caption{The probability that the full-sky and cut-sky minima (cold spot) are equivalent for any $\Delta\mathcal{T}_{\Psi}$ and for the CS's $\Delta\mathcal{T}_{\Psi}^{CS}$ is indicated by $\mathbb{P}(full)$ and $\mathbb{P}\big(full| \Delta\mathcal{T}_{\Psi}^{CS}\big)$, respectively, for each angular scale $R$. The significance of the CS is shown in comparison to the cut-sky and full-sky minima in CMB realisations. For each value of $R$ (except $R=5$ where 100,000 realisation were previously made) 10,000 CMB realisations were simulated.}
	\begin{tabular}{lcccc}
		\hline
		$R$ $[^{\circ}]$ & $\mathbb{P}(full)$ & $\mathbb{P}\left(full|\Delta\mathcal{T}_{\Psi}^{CS}\right)$ & Cut-sky $(\sigma)$ & Full-sky $(\sigma)$\\
		\hline
		$4$   & $0.51$ & $0.94$ & $1.95$ & $1.65$ \\
		$4.5$ & $0.50$ & $0.96$ & $2.18$ & $1.85$ \\
		$\boldsymbol{5}$   & $\boldsymbol{0.48}$ & $\boldsymbol{0.94}$ & $\boldsymbol{2.19}$ & $\boldsymbol{1.91}$ \\
		$5.5$ & $0.46$ & $0.96$ & $2.19$ & $1.89$ \\
		$6$   & $0.45$ & $0.94$ & $2.08$ & $1.76$ \\
		$6.5$ & $0.44$ & $0.91$ & $1.85$ & $1.50$ \\
		$7$   & $0.42$ & $0.86$ & $1.53$ & $1.13$ \\
		\hline
	\end{tabular}
	\label{table}
\end{table}

\subsection{Dependence on angular scale}

Up to this point we have used a preselected angular scale, $R=5^{\circ}$, where the CS was measured to be most significant by \citet{PlanckIso2015}. However, our conclusions for the CS significance may not necessarily hold true for other angular scales. To test this, $R$ is varied between $4^{\circ}$ and $7^{\circ}$, roughly equaling the range of $R$ over which \citet{Zhang2010} found the CS to be significant. The same procedure is carried out as before except with a smaller number of realisations (10,000).

In Table \ref{table} we summarise these results. The probability, $\mathbb{P}(full)$, is roughly equal to the fraction of unmasked pixels of the effective mask. However, $\mathbb{P}(full|\mathcal{T}_{\Psi}^{CS})$ is found to be $ > 0.85$ for the angular scales considered. When the cut-sky significance $> 2\sigma$ the probability is even higher ($> 0.93$). This makes it appropriate to compare the CS to full-sky minima in simulations where it is $< 2\sigma$ for $4^{\circ}<R<7^{\circ}$. Combined with previous studies \citep[e.g.][]{Vielva2004} means the CS is $< 2\sigma$ for all angular scales.

\section{Conclusions}

We measure the cut-sky and full-sky minima (cold spot) in 100,000 simulations using the Planck Common Field mask which has a similar $f_{sky}$ to the WMAP KQ75 and Planck U74 masks used in \citet{Zhang2010} and \citet{Nadathur2014} respectively. The probability of observing the full-sky minima is found to be $\sim 0.48$ (which roughly equals the unmasked fraction of the effective mask). At other positions the cut-sky minima is not equivalent to the full-sky minima and this biases the distribution of minima (see Fig. \ref{obsvstrue}). This appears to have a significant effect only at $\Delta \mathcal{T}_{\Psi}>-18\mu K$; at the CS's $\Delta \mathcal{T}_{\Psi} \simeq -19.3\mu K$ there is a $\sim 0.94$ probability that we are observing the CMB's full-sky minima.

We argue that the CS is detected as an anomaly, with a significance of $\sim2.2\sigma$, because the full-sky minimum is not always measured when using a mask resulting in an ensemble of cold spots which are slightly hotter than the full-sky ensemble. Correcting for this bias, by comparing to full-sky minima, reduces the significance to $\sim1.9\sigma$. We emphasize that the CS itself does not change due to the mask; rather, the ensemble to which it is compared is colder when the mask is removed. The difference in $\Delta T(\theta)$ and $\Delta T_{wav}(\theta)$ of the cut-sky and full-sky minima is subtle (see Fig. \ref{cstempprof}). But, a colder $\Delta T(\theta)$ for $\theta < 5^{\circ}$ results in colder $\Delta \mathcal{T}_{\Psi}$. This result is true for all angular scales (see Table \ref{table}) and would presumably remain for any model that can reproduce the CMB temperature $C_{\ell}$. In this sense these results are model independent. 

By varying the size of the mask, we find that the cut-sky minima is often not equal to the full-sky minima due to the latter's frequent obstruction by the mask. The inclusion of these hotter cut-sky minima appear to be driving the CS's significance. The CS can only be considered an anomaly if it is not the full-sky minimum itself as this would require a more extreme feature within the mask. This is unlikely, since such features are not seen in maps with a smaller mask or in full sky reconstructed maps \citep{Rassat2014}.

We investigate the ISW contribution (predicted by $\Lambda$CDM) to the coldest spots finding it to be poorly constrained and consistent with zero, but leaning towards a negative contribution (see Fig. \ref{isw}). On average it amounts to $\sim 10$ per cent of the full profile. Measurements of large voids in the LOS and ISW reconstructions are consistent with this result. Since reconstructed ISW profiles \citep[see][]{Nadathur2014,Finelli2015,AndrasJuan2015} appear to be below the mean shown in Fig. \ref{isw}, it is possible that the ISW is amplifying the significance of the CS. This would mean the primordial CS profile is even less significant than measured. Alternative models, which are not investigated here, may explain the slightly higher than expected causal relation between the observed and expected ISW of large voids seen in certain studies \citep{Granett2008,Cai2014,DES2017,Kovacs2017} but not all \citep{Ilic2013,Hotchkiss2015,NadaCrit2016}. Whether this is the case could be studied in future and would have implications for the predicted ISW contribution to the CS.

\section*{Acknowledgements}

We thank Andr\'{a}s Kovacs, Andrew Pontzen, Juan Garc\'{i}a-Bellido, Robert Crittenden, Seshadri Nadathur and Lorne Whiteway for providing suggestions which helped improve the paper. KN acknowledges support from the Science and Technology Facilities Council grant ST/N50449X. ABL thanks CNES for financial support through its post-doctoral programme. OL acknowledges support from a European Research Council Advanced Grant FP7/291329. Some of the results were derived using HEALPIX \citep{Healpix2005}.



\bibliographystyle{mnras}
\bibliography{bibfile}

\begin{thebibliography}{}
\makeatletter
\relax
\def\mn@urlcharsother{\let\do\@makeother \do\$\do\&\do\#\do\^\do\_\do\%\do\~}
\def\mn@doi{\begingroup\mn@urlcharsother \@ifnextchar [ {\mn@doi@}
  {\mn@doi@[]}}
\def\mn@doi@[#1]#2{\def\@tempa{#1}\ifx\@tempa\@empty \href
  {http://dx.doi.org/#2} {doi:#2}\else \href {http://dx.doi.org/#2} {#1}\fi
  \endgroup}
\def\mn@eprint#1#2{\mn@eprint@#1:#2::\@nil}
\def\mn@eprint@arXiv#1{\href {http://arxiv.org/abs/#1} {{\tt arXiv:#1}}}
\def\mn@eprint@dblp#1{\href {http://dblp.uni-trier.de/rec/bibtex/#1.xml}
  {dblp:#1}}
\def\mn@eprint@#1:#2:#3:#4\@nil{\def\@tempa {#1}\def\@tempb {#2}\def\@tempc
  {#3}\ifx \@tempc \@empty \let \@tempc \@tempb \let \@tempb \@tempa \fi \ifx
  \@tempb \@empty \def\@tempb {arXiv}\fi \@ifundefined
  {mn@eprint@\@tempb}{\@tempb:\@tempc}{\expandafter \expandafter \csname
  mn@eprint@\@tempb\endcsname \expandafter{\@tempc}}}

\bibitem[\protect\citeauthoryear{{Bennett} et~al.,}{{Bennett}
  et~al.}{2011}]{Wmap2011}
{Bennett} C.~L.,  et~al., 2011, \mn@doi [\apjs] {10.1088/0067-0049/192/2/17},
  \href {http://adsabs.harvard.edu/abs/2011ApJS..192...17B} {192, 17}

\bibitem[\protect\citeauthoryear{{Blas}, {Lesgourgues}  \& {Tram}}{{Blas}
  et~al.}{2011}]{CLASS}
{Blas} D.,  {Lesgourgues} J.,   {Tram} T.,  2011, \mn@doi [\jcap]
  {10.1088/1475-7516/2011/07/034}, \href
  {http://adsabs.harvard.edu/abs/2011JCAP...07..034B} {7, 034}

\bibitem[\protect\citeauthoryear{{Bobin}, {Sureau}, {Starck}, {Rassat}  \&
  {Paykari}}{{Bobin} et~al.}{2014}]{Bobin2014}
{Bobin} J.,  {Sureau} F.,  {Starck} J.-L.,  {Rassat} A.,   {Paykari} P.,  2014,
  \mn@doi [\aap] {10.1051/0004-6361/201322372}, \href
  {http://adsabs.harvard.edu/abs/2014A%26A...563A.105B} {563, A105}

\bibitem[\protect\citeauthoryear{{Bremer}, {Silk}, {Davies}  \&
  {Lehnert}}{{Bremer} et~al.}{2010}]{Bremer2010}
{Bremer} M.~N.,  {Silk} J.,  {Davies} L.~J.~M.,   {Lehnert} M.~D.,  2010,
  \mn@doi [\mnras] {10.1111/j.1745-3933.2010.00837.x}, \href
  {http://adsabs.harvard.edu/abs/2010MNRAS.404L..69B} {404, L69}

\bibitem[\protect\citeauthoryear{{Cai}, {Neyrinck}, {Szapudi}, {Cole}  \&
  {Frenk}}{{Cai} et~al.}{2014}]{Cai2014}
{Cai} Y.-C.,  {Neyrinck} M.~C.,  {Szapudi} I.,  {Cole} S.,   {Frenk} C.~S.,
  2014, \mn@doi [\apj] {10.1088/0004-637X/786/2/110}, \href
  {http://adsabs.harvard.edu/abs/2014ApJ...786..110C} {786, 110}

\bibitem[\protect\citeauthoryear{{Cay{\'o}n}, {Sanz},
  {Mart{\'{\i}}nez-Gonz{\'a}lez}, {Banday}, {Arg{\"u}eso}, {Gallegos},
  {G{\'o}rski}  \& {Hinshaw}}{{Cay{\'o}n} et~al.}{2001}]{Cayon2001}
{Cay{\'o}n} L.,  {Sanz} J.~L.,  {Mart{\'{\i}}nez-Gonz{\'a}lez} E.,  {Banday}
  A.~J.,  {Arg{\"u}eso} F.,  {Gallegos} J.~E.,  {G{\'o}rski} K.~M.,   {Hinshaw}
  G.,  2001, \mn@doi [\mnras] {10.1111/j.1365-8711.2001.04641.x}, \href
  {http://adsabs.harvard.edu/abs/2001MNRAS.326.1243C} {326, 1243}

\bibitem[\protect\citeauthoryear{{Cruz}, {Cay{\'o}n},
  {Mart{\'{\i}}nez-Gonz{\'a}lez}, {Vielva}  \& {Jin}}{{Cruz}
  et~al.}{2007}]{Wmap2007}
{Cruz} M.,  {Cay{\'o}n} L.,  {Mart{\'{\i}}nez-Gonz{\'a}lez} E.,  {Vielva} P.,
  {Jin} J.,  2007, \mn@doi [\apj] {10.1086/509703}, \href
  {http://adsabs.harvard.edu/abs/2007ApJ...655...11C} {655, 11}

\bibitem[\protect\citeauthoryear{{Finelli}, {Garc{\'{\i}}a-Bellido},
  {Kov{\'a}cs}, {Paci}  \& {Szapudi}}{{Finelli} et~al.}{2016}]{Finelli2015}
{Finelli} F.,  {Garc{\'{\i}}a-Bellido} J.,  {Kov{\'a}cs} A.,  {Paci} F.,
  {Szapudi} I.,  2016, \mn@doi [\mnras] {10.1093/mnras/stv2388}, \href
  {http://adsabs.harvard.edu/abs/2016MNRAS.455.1246F} {455, 1246}

\bibitem[\protect\citeauthoryear{{G{\'o}rski}, {Hivon}, {Banday}, {Wandelt},
  {Hansen}, {Reinecke}  \& {Bartelmann}}{{G{\'o}rski}
  et~al.}{2005}]{Healpix2005}
{G{\'o}rski} K.~M.,  {Hivon} E.,  {Banday} A.~J.,  {Wandelt} B.~D.,  {Hansen}
  F.~K.,  {Reinecke} M.,   {Bartelmann} M.,  2005, \mn@doi [\apj]
  {10.1086/427976}, \href {http://adsabs.harvard.edu/abs/2005ApJ...622..759G}
  {622, 759}

\bibitem[\protect\citeauthoryear{{Granett}, {Neyrinck}  \& {Szapudi}}{{Granett}
  et~al.}{2008}]{Granett2008}
{Granett} B.~R.,  {Neyrinck} M.~C.,   {Szapudi} I.,  2008, \mn@doi [\apjl]
  {10.1086/591670}, \href {http://adsabs.harvard.edu/abs/2008ApJ...683L..99G}
  {683, L99}

\bibitem[\protect\citeauthoryear{{Granett}, {Szapudi}  \& {Neyrinck}}{{Granett}
  et~al.}{2010}]{Granett2010}
{Granett} B.~R.,  {Szapudi} I.,   {Neyrinck} M.~C.,  2010, \mn@doi [\apj]
  {10.1088/0004-637X/714/1/825}, \href
  {http://adsabs.harvard.edu/abs/2010ApJ...714..825G} {714, 825}

\bibitem[\protect\citeauthoryear{{Hotchkiss}, {Nadathur}, {Gottl{\"o}ber},
  {Iliev}, {Knebe}, {Watson}  \& {Yepes}}{{Hotchkiss}
  et~al.}{2015}]{Hotchkiss2015}
{Hotchkiss} S.,  {Nadathur} S.,  {Gottl{\"o}ber} S.,  {Iliev} I.~T.,  {Knebe}
  A.,  {Watson} W.~A.,   {Yepes} G.,  2015, \mn@doi [\mnras]
  {10.1093/mnras/stu2072}, \href
  {http://adsabs.harvard.edu/abs/2015MNRAS.446.1321H} {446, 1321}

\bibitem[\protect\citeauthoryear{{Ili{\'c}}, {Langer}  \& {Douspis}}{{Ili{\'c}}
  et~al.}{2013}]{Ilic2013}
{Ili{\'c}} S.,  {Langer} M.,   {Douspis} M.,  2013, \mn@doi [\aap]
  {10.1051/0004-6361/201321150}, \href
  {http://adsabs.harvard.edu/abs/2013A%26A...556A..51I} {556, A51}

\bibitem[\protect\citeauthoryear{{Inoue} \& {Silk}}{{Inoue} \&
  {Silk}}{2006}]{Inoue12007}
{Inoue} K.~T.,  {Silk} J.,  2006, \mn@doi [\apj] {10.1086/505636}, \href
  {http://adsabs.harvard.edu/abs/2006ApJ...648...23I} {648, 23}

\bibitem[\protect\citeauthoryear{{Inoue} \& {Silk}}{{Inoue} \&
  {Silk}}{2007}]{Inoue22007}
{Inoue} K.~T.,  {Silk} J.,  2007, \mn@doi [\apj] {10.1086/517603}, \href
  {http://adsabs.harvard.edu/abs/2007ApJ...664..650I} {664, 650}

\bibitem[\protect\citeauthoryear{{Kov{\'a}cs}}{{Kov{\'a}cs}}{2017}]{Kovacs2017}
{Kov{\'a}cs} A.,  2017, preprint, \href
  {http://adsabs.harvard.edu/abs/2017arXiv170108583K} {} (\mn@eprint {arXiv}
  {1701.08583})

\bibitem[\protect\citeauthoryear{{Kov{\'a}cs} \&
  {Garc{\'{\i}}a-Bellido}}{{Kov{\'a}cs} \&
  {Garc{\'{\i}}a-Bellido}}{2016}]{AndrasJuan2015}
{Kov{\'a}cs} A.,  {Garc{\'{\i}}a-Bellido} J.,  2016, \mn@doi [\mnras]
  {10.1093/mnras/stw1752}, \href
  {http://adsabs.harvard.edu/abs/2016MNRAS.462.1882K} {462, 1882}

\bibitem[\protect\citeauthoryear{{Kov{\'a}cs} et~al.,}{{Kov{\'a}cs}
  et~al.}{2017}]{DES2017}
{Kov{\'a}cs} A.,  et~al., 2017, \mn@doi [\mnras] {10.1093/mnras/stw2968}, \href
  {http://adsabs.harvard.edu/abs/2017MNRAS.465.4166K} {465, 4166}

\bibitem[\protect\citeauthoryear{{Mackenzie}, {Shanks}, {Bremer}, {Cai},
  {Gunawardhana}, {Kov{\'a}cs}, {Norberg}  \& {Szapudi}}{{Mackenzie}
  et~al.}{2017}]{Mackenzie2017}
{Mackenzie} R.,  {Shanks} T.,  {Bremer} M.~N.,  {Cai} Y.-C.,  {Gunawardhana}
  M.~L.~P.,  {Kov{\'a}cs} A.,  {Norberg} P.,   {Szapudi} I.,  2017, preprint,
  \href {http://adsabs.harvard.edu/abs/2017arXiv170403814M} {} (\mn@eprint
  {arXiv} {1704.03814})

\bibitem[\protect\citeauthoryear{{Marcos-Caballero}, {Fern{\'a}ndez-Cobos},
  {Mart{\'{\i}}nez-Gonz{\'a}lez}  \& {Vielva}}{{Marcos-Caballero}
  et~al.}{2016}]{Marcos2016}
{Marcos-Caballero} A.,  {Fern{\'a}ndez-Cobos} R.,
  {Mart{\'{\i}}nez-Gonz{\'a}lez} E.,   {Vielva} P.,  2016, \mn@doi [\mnras]
  {10.1093/mnrasl/slw063}, \href
  {http://adsabs.harvard.edu/abs/2016MNRAS.460L..15M} {460, L15}

\bibitem[\protect\citeauthoryear{{Marcos-Caballero},
  {Mart{\'{\i}}nez-Gonz{\'a}lez}  \& {Vielva}}{{Marcos-Caballero}
  et~al.}{2017}]{Marcos2017}
{Marcos-Caballero} A.,  {Mart{\'{\i}}nez-Gonz{\'a}lez} E.,   {Vielva} P.,
  2017, \mn@doi [\jcap] {10.1088/1475-7516/2017/02/026}, \href
  {http://adsabs.harvard.edu/abs/2017JCAP...02..026M} {2, 026}

\bibitem[\protect\citeauthoryear{{Nadathur} \& {Crittenden}}{{Nadathur} \&
  {Crittenden}}{2016}]{NadaCrit2016}
{Nadathur} S.,  {Crittenden} R.,  2016, \mn@doi [\apjl]
  {10.3847/2041-8205/830/1/L19}, \href
  {http://adsabs.harvard.edu/abs/2016ApJ...830L..19N} {830, L19}

\bibitem[\protect\citeauthoryear{{Nadathur}, {Lavinto}, {Hotchkiss}  \&
  {R{\"a}s{\"a}nen}}{{Nadathur} et~al.}{2014}]{Nadathur2014}
{Nadathur} S.,  {Lavinto} M.,  {Hotchkiss} S.,   {R{\"a}s{\"a}nen} S.,  2014,
  \mn@doi [\prd] {10.1103/PhysRevD.90.103510}, \href
  {http://adsabs.harvard.edu/abs/2014PhRvD..90j3510N} {90, 103510}

\bibitem[\protect\citeauthoryear{{Naidoo}, {Benoit-L{\'e}vy}  \&
  {Lahav}}{{Naidoo} et~al.}{2016}]{Naidoo2016}
{Naidoo} K.,  {Benoit-L{\'e}vy} A.,   {Lahav} O.,  2016, \mn@doi [\mnras]
  {10.1093/mnrasl/slw043}, \href
  {http://adsabs.harvard.edu/abs/2016MNRAS.459L..71N} {459, L71}

\bibitem[\protect\citeauthoryear{{Planck Collaboration} et~al.,}{{Planck
  Collaboration} et~al.}{2016a}]{PlanckPara2015}
{Planck Collaboration} et~al., 2016a, \mn@doi [\aap]
  {10.1051/0004-6361/201525830}, \href
  {http://adsabs.harvard.edu/abs/2016A%26A...594A..13P} {594, A13}

\bibitem[\protect\citeauthoryear{{Planck Collaboration} et~al.,}{{Planck
  Collaboration} et~al.}{2016b}]{PlanckIso2015}
{Planck Collaboration} et~al., 2016b, \mn@doi [\aap]
  {10.1051/0004-6361/201526681}, \href
  {http://adsabs.harvard.edu/abs/2016A%26A...594A..16P} {594, A16}

\bibitem[\protect\citeauthoryear{{Planck Collaboration} et~al.,}{{Planck
  Collaboration} et~al.}{2016c}]{PlanckISW2016}
{Planck Collaboration} et~al., 2016c, \mn@doi [\aap]
  {10.1051/0004-6361/201525831}, \href
  {http://adsabs.harvard.edu/abs/2016A%26A...594A..21P} {594, A21}

\bibitem[\protect\citeauthoryear{{Rassat}, {Starck}, {Paykari}, {Sureau}  \&
  {Bobin}}{{Rassat} et~al.}{2014}]{Rassat2014}
{Rassat} A.,  {Starck} J.-L.,  {Paykari} P.,  {Sureau} F.,   {Bobin} J.,  2014,
  \mn@doi [\jcap] {10.1088/1475-7516/2014/08/006}, \href
  {http://adsabs.harvard.edu/abs/2014JCAP...08..006R} {8, 006}

\bibitem[\protect\citeauthoryear{{Rees} \& {Sciama}}{{Rees} \&
  {Sciama}}{1968}]{Rees1968}
{Rees} M.~J.,  {Sciama} D.~W.,  1968, \mn@doi [\nat] {10.1038/217511a0}, \href
  {http://adsabs.harvard.edu/abs/1968Natur.217..511R} {217, 511}

\bibitem[\protect\citeauthoryear{{Rogers}, {Peiris}, {Leistedt}, {McEwen}  \&
  {Pontzen}}{{Rogers} et~al.}{2016}]{Rogers2016}
{Rogers} K.~K.,  {Peiris} H.~V.,  {Leistedt} B.,  {McEwen} J.~D.,   {Pontzen}
  A.,  2016, \mn@doi [\mnras] {10.1093/mnras/stw1121}, \href
  {http://adsabs.harvard.edu/abs/2016MNRAS.460.3014R} {460, 3014}

\bibitem[\protect\citeauthoryear{{Sachs} \& {Wolfe}}{{Sachs} \&
  {Wolfe}}{1967}]{Sachs1967}
{Sachs} R.~K.,  {Wolfe} A.~M.,  1967, \mn@doi [\apj] {10.1086/148982}, \href
  {http://adsabs.harvard.edu/abs/1967ApJ...147...73S} {147, 73}

\bibitem[\protect\citeauthoryear{{Szapudi} et~al.,}{{Szapudi}
  et~al.}{2015}]{Szapudi2015}
{Szapudi} I.,  et~al., 2015, \mn@doi [\mnras] {10.1093/mnras/stv488}, \href
  {http://adsabs.harvard.edu/abs/2015MNRAS.450..288S} {450, 288}

\bibitem[\protect\citeauthoryear{{Vielva}, {Mart{\'{\i}}nez-Gonz{\'a}lez},
  {Barreiro}, {Sanz}  \& {Cay{\'o}n}}{{Vielva} et~al.}{2004}]{Vielva2004}
{Vielva} P.,  {Mart{\'{\i}}nez-Gonz{\'a}lez} E.,  {Barreiro} R.~B.,  {Sanz}
  J.~L.,   {Cay{\'o}n} L.,  2004, \mn@doi [\apj] {10.1086/421007}, \href
  {http://adsabs.harvard.edu/abs/2004ApJ...609...22V} {609, 22}

\bibitem[\protect\citeauthoryear{{Zhang} \& {Huterer}}{{Zhang} \&
  {Huterer}}{2010}]{Zhang2010}
{Zhang} R.,  {Huterer} D.,  2010, \mn@doi [Astroparticle Physics]
  {10.1016/j.astropartphys.2009.11.005}, \href
  {http://adsabs.harvard.edu/abs/2010APh....33...69Z} {33, 69}

\bibitem[\protect\citeauthoryear{{Zibin}}{{Zibin}}{2014}]{Zibin2014}
{Zibin} J.~P.,  2014, preprint, \href
  {http://adsabs.harvard.edu/abs/2014arXiv1408.4442Z} {} (\mn@eprint {arXiv}
  {1408.4442})

\makeatother
\end{thebibliography}




\label{lastpage}
\end{document}